\documentclass[12pt]{article}

\usepackage{amssymb}
\usepackage{amsmath}

\usepackage{graphicx}

\begin{document} %
%%%%%%%%%%%%%%%%%%

%%%%%%%%%%%%%%%%%%%%%%%%%%%%%%%%%%%%%%%%%%%%%%%%%%%%%%%%%%%%%%%%%%%%%

\title{A search for anomalous quartic gauge couplings in lead-lead collisions at the FCC-hh}

\author{S.C. \.{I}nan\thanks{Electronic address: sceminan@cumhuriyet.edu.tr}
\\
{\small Department of Physics, Sivas Cumhuriyet University, 58140,
Sivas, Turkey}
\\
{\small and}
\\
A.V. Kisselev\thanks{Electronic address:
alexandre.kisselev@ihep.ru} \\
{\small A.A. Logunov Institute for High Energy Physics, NRC
``Kurchatov Institute'',}
\\
{\small 142281, Protvino, Russian Federation} }

\date{}

\maketitle

\begin{abstract}
A sensitivity of the collider FCC-hh to anomalous quartic gauge
couplings (aQGCs) of neutral bosons in lead-lead collisions is
studied. Examining diphoton production, the bounds on aQGCs at
$5\sigma$ and 95\% C.L. are obtained depending on systematic
uncertainties.
\end{abstract}

\maketitle

%%%%%%%%%%%%%%%%%%%%%%%%%%%%%%%%%%%%%%%%%%%%%%%%%%%%%%%%%%%%%%%%%%%%%

%%%%%%%%%%%%%%%%%%%%%%%%
\section{Introduction} %
%%%%%%%%%%%%%%%%%%%%%%%%
In a number of our previous papers, we examined anomalous gauge
coupling constants (aQGCs) in collisions at the CLIC
\cite{I_K:2021_1,I_K:2021_2} and muon collider
\cite{I_K:2024,I_K:2025}. As a result, exclusion bounds on neutral
aQGCs we were derived in \cite{I_K:2021_1}-\cite{I_K:2025}. A search
for aQGCs was done in experiments at the LHC
\cite{ATLAS:QGCs_1}-\cite{CMS_TOTEM:QGCs}. The aQGCs with neutral
gauge bosons are of particular interest, since such vertices are
absent in the Lagrangian of the Standard Model (SM).
Phenomenological bounds on neutral aQGCs were obtained both for the
CLIC \cite{I_K:2021_1,I_K:2021_2},
\cite{{Koksal:2014}}-\cite{Gurkanli:2024} and LHC
\cite{{Eboli:2004}}-\cite{Royon:2024}. The upper bounds on aQGCs
were also found for future $\gamma e$ and $\gamma\gamma$ colliders
\cite{Eboli:1994}-\cite{Koksal:2016}, future hadron
\cite{Senol:2022_1}-\cite{Senol:2022}, and $eh$ colliders
\cite{Gurkanli:2020}-\cite{Gurkanli:2023_1}. Recently expected
limits on aQGCs at a muon collider have been examined
\cite{I_K:2024,I_K:2025}, \cite{Abbott:2021}-\cite{Zhang:2023}.

The coherent action of $Z$ protons in a heavy-ion collision provides
an intensive spectrum of equivalent photons in the ions with nuclear
charge $Z$. Since each photon flux scales as the square of the ion
charge $Z^2$, the $\gamma\gamma$ process in lead-lead (Pb + Pb)
collision, which occurs at one-loop level at order
$\alpha_{\mathrm{em}}^4$ and has a tiny cross section, are enhanced
by a factor of $Z^4 \simeq 5 \times 10^7$ compared to similar $pp$
or $e^+e^-$ interactions. As a consequence, in ultraperipheral heavy
ion collisions, where the impact parameter is larger than the radius
and the ions scatter quasi-elastically, the elementary
$\gamma\gamma\rightarrow\gamma\gamma$ process, which occurs at
one-loop level at order $\alpha_{\mathrm{em}}^4$ and have a tiny
cross section, is enhanced by a large $Z^4$ ($4.5\times10^7$)
factor. In addition, the contribution of gluon initiated processes
can be strongly reduced in nuclear collisions \cite{Cuelho:2020},
becoming the light-by-light (LbL) scattering feasible for the
experimental analysis.

Thus, the coherent action of $Z$ protons in the heavy-ion collision
provides an intensive spectrum of quasi-real photons in the ions.
Then the cross section for the elastic process $\mathrm{Pb}
\,\mathrm{Pb} \rightarrow \mathrm{Pb} \,\gamma\gamma\, \mathrm{Pb}$
can be calculated by convolving the appropriate photon flux with the
elementary cross section for the process $\gamma\gamma \rightarrow
\gamma\gamma$. The LbL scattering in heavy-ion collisions at the LHC
was observed in \cite{ATLAS_ions_1}-\cite{CMS_ions_3}. The cross
sections are consistent with the theoretical prediction at NLO
accuracy in QCD and QED \cite{Chaubey:2024_1,Chaubey:2024_2}.

In the present paper, we study aQGCs by considering photon pair
production in lead-lead collisions at the 100 TeV FCC-hh
\cite{FCC_1,FCC_2}. Details of a physics potential of heavy-ion
running at the FCC-hh are provided in
refs.~\cite{Mangano:2017,Doinese:2017}. To examine aQGCs we use the
Effective Field Theory (EFT) model-independent framework. The
effective Lagrangian of dimension-8 operators describing aQGCs is
given by \cite{Gurkanli:2024,Degrande:2013, Almeida:2020}
\begin{equation}\label{EFT_Lagragian}
\mathcal{L}_{\mathrm{eff}} = \mathcal{L}_{\mathrm{SM}} +
\sum_{i=0}^1 \frac{f_{S,i}}{\Lambda^4}\,\mathcal{O}_{S,i} +
\sum_{i=0}^7 \frac{f_{M,i}}{\Lambda^4}\,\mathcal{O}_{M,i} +
\sum_{\substack{i=0 \\ i\neq 3,4}}^9
\frac{f_{T,i}}{\Lambda^4}\,\mathcal{O}_{T,i} \;.
\end{equation}
There are three classes of three types of dimension-8 operators in
this Lagrangian: i) operators $O_{S,i}$ containing covariant
derivatives of the Higgs doublet only; ii) operators $O_{M,i}$ with
two field strength tensors and two derivatives of Higgs; iii)
operators  $O_{T,i}$ with field strength tensors only. It is the
tensor operators $O_{T,i}$  which induce anomalous
$\gamma\gamma\gamma\gamma$ vertex. They are as follows (see, for
instance, \cite{Gurkanli:2024})
\begin{align}\label{dim8_operators}
O_{T,0} &= \frac{f_{T,0}}{\Lambda^4} \mathrm{Tr}\left[ W_{\mu\nu}
W^{\mu\nu}\right] \times \mathrm{Tr}\left[ W_{\alpha\beta}
W^{\alpha\beta}\right] , \nonumber \\
O_{T,1} &= \frac{f_{T,1}}{\Lambda^4} \mathrm{Tr}\left[ W_{\alpha\mu}
W^{\nu\beta}\right] \times \mathrm{Tr}\left[ W_{\nu\beta}
W^{\alpha\mu}\right] , \nonumber \\
O_{T,2} &= \frac{f_{T,2}}{\Lambda^4} \mathrm{Tr}\left[ W_{\alpha\mu}
W^{\mu\beta}\right] \times \mathrm{Tr}\left[ W_{\beta\nu}
W^{\nu\alpha}\right] , \nonumber \\
O_{T,5} &= \frac{f_{T,5}}{\Lambda^4} \mathrm{Tr}\left[ W_{\mu\nu}
W^{\mu\nu}\right] \times B_{\alpha\beta} B^{\alpha\beta} , \nonumber \\
O_{T,6} &= \frac{f_{T,6}}{\Lambda^4} \mathrm{Tr}\left[ W_{\alpha\mu}
W^{\nu\beta}\right] \times B_{\nu\beta} B^{\alpha\mu} , \nonumber \\
O_{T,7} &= \frac{f_{T,7}}{\Lambda^4} \mathrm{Tr}\left[ W_{\alpha\mu}
W^{\mu\beta}\right] \times B_{\beta\nu} B^{\nu\alpha} , \nonumber \\
O_{T,8} &= \frac{f_{T,8}}{\Lambda^4} B_{\mu\nu} B^{\mu\nu} \times
B_{\alpha\beta} B^{\alpha\beta} , \nonumber \\
O_{T,9} &= \frac{f_{T,9}}{\Lambda^4} B_{\alpha\mu} B^{\mu\beta}
\times B_{\beta\nu} B^{\nu\alpha} .
\end{align}
Here $f_{T,i}$ ($i = 0,1,2,5,6,7,8,9$) are dimensionless parameters,
and $\Lambda$ is a mass-dimension parameter associated with the
scale of physics beyond the SM. Note that dimension-8 operators are
lowest-dimension operators inducing only aQGCs without anomalous
triple gauge boson vertices.

After EW symmetry breaking the part of the Lagrangian describing
$\gamma\gamma\gamma\gamma$ interaction is defined by two operators,
\begin{equation}\label{4photon_Lagrangian}
\mathcal{L}_{\gamma\gamma\gamma\gamma} = g_1 F_{\mu\nu}F^{\mu\nu}
F_{\alpha\beta}F^{\alpha\beta}  + g_2 F_{\mu\nu}F^{\nu\alpha}
F_{\alpha\beta}F^{\beta\mu} \;,
\end{equation}
with two anomalous couplings $g_1$ and $g_2$ of dimension -4. They
are linear combinations of the ``unbroken'' couplings
$f_{T,i}/\Lambda^4$,
\begin{align}\label{g1_g2_vs_fTi}
g_1 &= \frac{s_w^4}{\Lambda^4} (f_{T,0} + f_{T,1})
+ \frac{s_w^2 c_w^2}{\Lambda^4} (f_{T,5}+ f_{T,6}) + \frac{c_w^4}{\Lambda^4} f_{T,8} \;,
\nonumber \\
g_2 &= \frac{s_w^4}{\Lambda^4} f_{T,2} + \frac{s_w^2
c_w^2}{\Lambda^4}f_{T,7} + \frac{c_w^4}{\Lambda^4} f_{T,9} \;.
\end{align}
As one can see from \eqref{g1_g2_vs_fTi}, the diphoton production
via the $\gamma\gamma\rightarrow\gamma\gamma$ scattering is more
sensitive to the anomalous couplings $f_{T,8}/\Lambda^4$ and
$f_{T,9}/\Lambda^4$.

The $\gamma\gamma\gamma\gamma$ couplings can be modified by loops of
new heavy charged particles, then we get an estimate ($i = 1,2$)
\begin{equation}\label{new_charged_part}
g_i \sim \alpha_{\mathrm{em}}^2 Q^2 m_{ch}^{-4} \;,
\end{equation}
with $Q$ and $m_{ch}$ being the charge and mass of this particle.
One can find that $g_i \sim 10^{-2}\div 10^{-1}$ TeV$^{-4}$. The
anomalous interaction of the neutral bosons can be also defined by
$s$-channel contribution from new (pseudo)scalar or spin-2 heavy
particle $X$. Then
\begin{equation}\label{new_charged_part}
g_i \sim (f_s m_s)^{-2} ,
\end{equation}
where $f_s$ is the $X\gamma\gamma$-coupling, and $m_s$ is its mass.
For instance, $X$ can be 2 TeV dilaton that leads to $g_i \sim
10^{-1}$ TeV$^{-4}$ \cite{Royon:2024}. If $X$ is the lightest
KK-graviton of the Randall-Sundrum model \cite{Randall:1999}, then
$f_s, m_s \sim$ few TeV, and $g_i$ is of order $10^{-2}$ TeV$^{-4}$.
If $X$ is a heavy axion-like particle (ALP) then
\begin{equation}\label{ALP_as_X}
g_i \sim g_{a\gamma\gamma}^2/m_a^2 \;,
\end{equation}
where $g_{a\gamma\gamma}$ is the ALP-photon coupling and $m_a$ is
the ALP mass. Taking the LHC bound $g_{a\gamma\gamma} \simeq 5\times
10^{-2}$ TeV$^{-1}$ for $m_a \simeq 1$ TeV
\cite{Esterria:2021,Hare:2020}, we get an estimate $g_i \sim
10^{-3}$ TeV$^{-4}$. To conclude, one may expect that the neutral
aQGCs can be of order $10^{-3}\div 10^{-1}$ TeV$^{-4}$ or smaller.

%%%%%%%%%%%%%%%%%%%%%%%%%%%%%%%%%%%%%%%%%%%%%%%%%%%%%%%%%%%%%%%%%%%%%%%%%%
\section{Diphoton production in nucleus-nucleus collision at the FCC-hh} %
%%%%%%%%%%%%%%%%%%%%%%%%%%%%%%%%%%%%%%%%%%%%%%%%%%%%%%%%%%%%%%%%%%%%%%%%%%

The final-state signature of the lead-lead collision is the
exclusive production of two photons,
\begin{equation}\label{process}
\mathrm{Pb} \,\mathrm{Pb} \rightarrow \mathrm{Pb} \,\gamma\gamma
\,\mathrm{Pb} \;,
\end{equation}
with the diphoton final-state measured in the central detector, and
lead nuclei surviving the electromagnetic interaction scattered at
very low angles with respect to the beams. In the equivalent photon
approximation (EPA) \cite{Budnev:1975} the accelerated lead ions can
be considered as $\gamma$ beams. Indeed, the emitted photons are
almost on-shell, since their virtuality $|Q^2| < 1/R_A^2$, where
$R_A = 1.2A^{1/3}$ fm is the radius of the nucleus. It results in
$|Q^2| < 7.7 \cdot 10^{-4}$ GeV$^2$ for A = 208.

In EPA the differential cross section of a photon fusion process $NN
\rightarrow N \,\gamma\gamma \, N$ can be factorized as
\begin{equation}\label{cs}
d\sigma = \int\limits_{\tau_{\min}}^{\tau_{\max}}
\!\!\frac{d\tau}{\tau}
\!\!\int\limits_{\omega_{\min}}^{\omega_{\max}}
\!\!\frac{d\omega}{\omega} f_{\gamma/N}(\omega)
f_{\gamma/N}(\tau/\omega) \,d\hat{\sigma} (\gamma\gamma\rightarrow
\gamma\gamma) \;,
\end{equation}
where
\begin{equation}\label{upper_limits}
\omega_{\max} = E_N - m_N \;, \quad \tau_{\max} = (E_N - m_N)^2 \;,
\end{equation}
$\omega$ is the photon energies emitted from the nucleus, $E_N$ is
the energy of the ion beams, and $m_N$ is the nucleus mass. The
quantity $4\tau$ coincides with the center-of-mass energy squared of
the process $\gamma\gamma\rightarrow\gamma\gamma$. As for the lower
limits on variables $\omega$ and $\tau$ in \eqref{cs}, they are
given by equations
\begin{equation}\label{lower_limits}
\omega_{\min} =  \tau/\omega_{\max} \;, \quad \tau_{\min} = p_\bot^2
\;,
\end{equation}
where $p_\bot$ is the transverse momenta of the outgoing photons.

In the relativistic limit the equivalent spectrum of the photon from
the nucleus $N$ with the charge $Z$ and atomic number $A$ is given
by \cite{Jackson:QED}-\cite{Baltz:2008} (see also
\cite{Baur:1990_1}-\cite{Akiba:2016}).
\begin{equation}\label{dist_gamma_nucleus}
f_{\gamma/N}(\omega) = \frac{2Z^2\alpha}{\pi} \!\left[ \xi K_0(\xi)
K_1(\xi) - \frac{\xi^2}{2} ( K_1^2(\xi) - K_0^2(\xi) ) \right] ,
\end{equation}
where $\xi = \omega/E_R$, $E_R = E_N/(m_N R_A) = \sqrt{s_{NN}}/(2m_p
R_A)$, $m_p$ being the nucleon mass. $K_0(x)$ ($K_1(x)$) is the
modified Bessel function of the second kind of order zero (one).
Note that
\begin{equation}\label{Bessels_appr}
\left[ \xi K_0(\xi) K_1(\xi) - \frac{\xi^2}{2} ( K_1^2(\xi) -
K_0^2(\xi) ) \right] \bigg|_{\xi \rightarrow 0} = \ln\frac{1}{\xi} -
C + \mathrm{O}(\xi^2) \;,
\end{equation}
where $C = \gamma_E + 0.5 - \ln2$, and $\gamma_E \simeq 0.5772$ is
Euler's constant, that leads to the approximation
\begin{equation}\label{dist_gamma_small_xi}
f_{\gamma/N}(x)|_{x \rightarrow 0} = \frac{Z^2\alpha}{\pi} \!\left[
\ln \frac{1}{(x m_N R_A)^2} - 0.768 \right] ,
\end{equation}
where a dimensionless variable $x = \omega/E_N$ is introduced. For
large values of $\xi$ we get
\begin{equation}\label{dist_gamma_large_xi}
\left[ \xi K_0(\xi) K_1(\xi) - \frac{\xi^2}{2} ( K_1^2(\xi) -
K_0^2(\xi) ) \right] \bigg|_{\xi \rightarrow \infty} = \frac{\pi}{4}
\,e^{-2\xi} \left[ 1 + \mathrm{O}(\xi^{-1}) \right].
\end{equation}

For the PbPb collisions (with $Z = 82$ and $A = 208$) at the 100 TeV
FCC-hh \cite{FCC_1,FCC_2} we find that that the beam momenta $p_N =
(\sqrt{s}/2)\,(Z/A)$ is equal to 19.7 TeV, the beam energy is 4100
TeV, and the energy per nucleon is $\sqrt{s_{NN}} = \sqrt{s}\,(Z/A)
= 39.4$ TeV. We also get
\begin{equation}\label{E_N}
R_A = 7.1 \mathrm{\ fm} = 36.0 \mathrm{\ GeV}^{-1} \;,  \ E_N
=4097.6 \mathrm{\ TeV} \;, \ E_R \simeq 582.8 \mathrm{\ GeV} \;,
\end{equation}
and
\begin{equation}\label{xi}
\xi \simeq 1.71 \left( \frac{\omega}{\mathrm{TeV}} \right) .
\end{equation}
The Lorentz factor is defined to be $\gamma_L = E_N/m_N = 20935.7$.
As one can derive from \eqref{dist_gamma_nucleus}, the photon
spectra have a $\omega^{-1}$ fall-off up to energies of the order of
$\omega_{\max} \thickapprox \gamma_L/R_A$ \cite{Enterria:2013}. In
our case, $\omega_{\max} \thickapprox 580$ GeV.

The exact helicity amplitudes of the
$\gamma\gamma\rightarrow\gamma\gamma$ collision are given in
Appendix~A. In a one-loop approximation, the SM contribution to the
cross section, coming from VBF process is a sum of the fermion and
$W$ boson amplitudes,
\begin{equation}\label{f+W_ew}
M^{\mathrm{SM}} = M^{\mathrm{SM}}_f + M^{\mathrm{SM}}_W  \;.
\end{equation}
The expressions for the SM helicity amplitudes $M^{\mathrm{SM}}_f$
and $M^{\mathrm{SM}}_W$ are known in explicit forms
\cite{Jikia:1994}-\cite{Gounaris:1999_3} that enables one to
calculate interference terms analytically. Note that
$M^{\mathrm{SM}}_W$ dominates over $M^{\mathrm{SM}}_f$ if
center-of-mass energy exceeds 200 GeV. As shown in
\cite{Bern:2001,Binoth:2002}, two-loop QCD and QED corrections to
LbL cross section by fermion loops in the ultrarelativistic limit,
where all kinematic invariants are much greater than the relevant
fermion masses, are quite small numerically (a few percent), showing
that the leading order computations are robust.
\footnote{The computation of NLO QCD and QED contributions to the
LbL cross-section, retaining exact dependence in the fermion masses,
is presented in refs.~\cite{Ajjath:2024_1,Ajjath:2024_2}.}

\begin{figure}[htb]
\begin{center}
\includegraphics[scale=0.52]{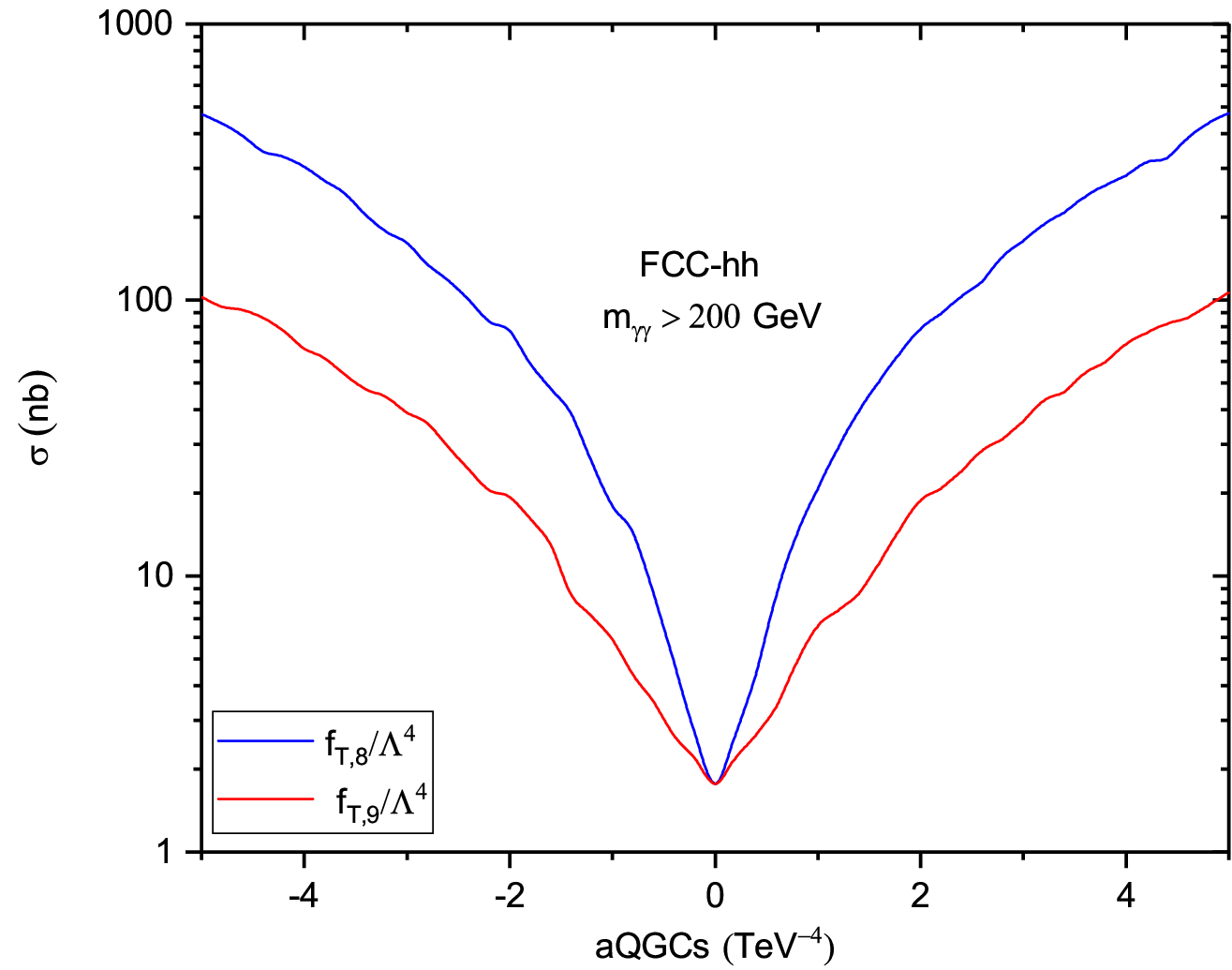}
\caption{The total cross section of the $\mathrm{Pb} \,\mathrm{Pb}
\rightarrow \mathrm{Pb}\, \gamma\gamma \,\mathrm{Pb}$ process as a
function of aQGCs $f_{T,8}/\Lambda^4$ (upper curve) and
$f_{T,9}/\Lambda^4$ (lower curve) at the FCC-hh.}
\label{fig:FCC_CS_ft8_ft9}
\end{center}
\end{figure}

The total cross section at the FCC-hh, depending on aQGC is given in
Fig.~\ref{fig:FCC_CS_ft8_ft9}. To compare our results with similar
results obtained recently for $pp$ collisions at the 14 TeV LHC
\cite{Senol:2025}, we focus on the two aQGCs, $f_{T,8}/\Lambda^4$
and $f_{T,9}/\Lambda^4$. In our calculations, we used the cut on the
photon rapidity of the outgoing photons, $|\eta^\gamma| < 2.5$. We
also imposed the cut on the photon transverse momenta $p_t^\gamma >
5$ GeV. Then the photon particle-identification efficiency is closed
to unity \cite{ATLAS_ions_2}. To additionally reduce the SM
background, we applied the cut on the invariant mass of the diphoton
system, $m_{\gamma\gamma} > 200$ GeV ($m_{\gamma\gamma} > 500$ GeV).

Note that the diphoton system is produced from quasi-real photons
almost at rest. That is why we impose very tight cut on the
transverse momentum of the pair momentum, $p_\bot^{\gamma\gamma} <
0.1$ GeV. To reduce a prompt-photon background from the central
exclusive process, $gg \rightarrow \gamma\gamma$, a requirement on
the acomplanarity, $|\Delta\phi - \pi| < 0.04$, is used.

An important background in ultraperipheral heavy-ion collisions can
come from exclusive $\gamma\gamma \rightarrow e^+e^-$ events
\cite{ATLAS:PbPb}-\cite{Enterria:2013}. The final-state electrons can
be misidentified as photons. Experimental study indicates that for
the heavy-ion collisions at the LHC, single-electron
misidentification probability is about $P_{e\rightarrow\gamma}
\approx (0.5-1)\%$ \cite{ATLAS:PbPb}-\cite{CMS_misident}. We assume
that the same $P_{e\rightarrow\gamma}$ is realized for the FCC-hh.
We have estimated the $\gamma\gamma \rightarrow e^+e^-$ cross
section to be 14.6 nb at $m_{\gamma\gamma} > 200$ GeV. Requiring
both $e^+$ and $e^-$ are mis-tagged results in a suppression of this
background by the factor $P_{e\rightarrow\gamma}^2 \approx (0.25 -
1)\cdot10^{-4}$. Thus, a rate of the $\gamma\gamma \rightarrow
e^+e^-$ background can be safely neglected.

Note that QED processes such as $\gamma\gamma \rightarrow
\mu^+\mu^-, \tau^+\tau^-, q\bar{q}$ are much smaller as their
final-states include charged particles in the tracker and/or muon
spectrometer \cite{Enterria:2013}. The hard bremsstrahlung photons
emitted in the $\gamma\gamma \rightarrow e^+e^- \gamma\gamma$
process, as estimated in \cite{ATLAS:PbPb}, to be below 1\% of an
expected signal (see also \cite{Enterria:2013}). The bremsstrahlung
photons from the ions can be ignored, since they have $p_\bot
\lesssim 1/R_A$, with  $R_A \simeq 36$ GeV$^{-1}$
\cite{Knapen:2017}. There is another potential backgrounds to the
LbL scattering in $\mathrm{Pb Pb}$ collisions, the double
diffractive process. But it is suppressed, if the cuts on
$p_t^{\gamma\gamma}$ and $|\eta^\gamma|$ are used
\cite{Cuelho:2020}. The contributions from meson decays into photons
\cite{Klusek:2019} become negligible after imposing strong lower cut
on $m_{\gamma\gamma}$ (in combination with other our cuts).
%
%%%%%%%%%%%%%%%%%%%%%%%%%%%%%%%%%%%%%%%%%%%%%%%%%%%%%
% Figure 1. Total cross section vs. aQGCs at FCC-hh %
%%%%%%%%%%%%%%%%%%%%%%%%%%%%%%%%%%%%%%%%%%%%%%%%%%%%%

In the next Fig.~\ref{fig:CSMCUT} the total cross section is
presented as a function of the minimal value of the diphoton
invariant mass $m_{\gamma\gamma}$. There are two curves
corresponding to fixed value of one of aQGCs ($f_{T,8}/\Lambda^4$ or
$f_{T,9}/\Lambda^4$). As one can see, the total cross section is
almost constant at $m_{\gamma\gamma}
> 250$ GeV, while the SM cross sections decrease very rapidly in
this region.
%
%%%%%%%%%%%%%%%%%%%%%%%%%%%%%%%%%%%%%%%%%%%%%%%%%%%%%%%%%%%%%%%%%%%%%%%
% Figure 2. Total cross section vs. diphoton invariant mass at FCC-hh %
%%%%%%%%%%%%%%%%%%%%%%%%%%%%%%%%%%%%%%%%%%%%%%%%%%%%%%%%%%%%%%%%%%%%%%%
\begin{figure}[htb]
\begin{center}
\includegraphics[scale=0.52]{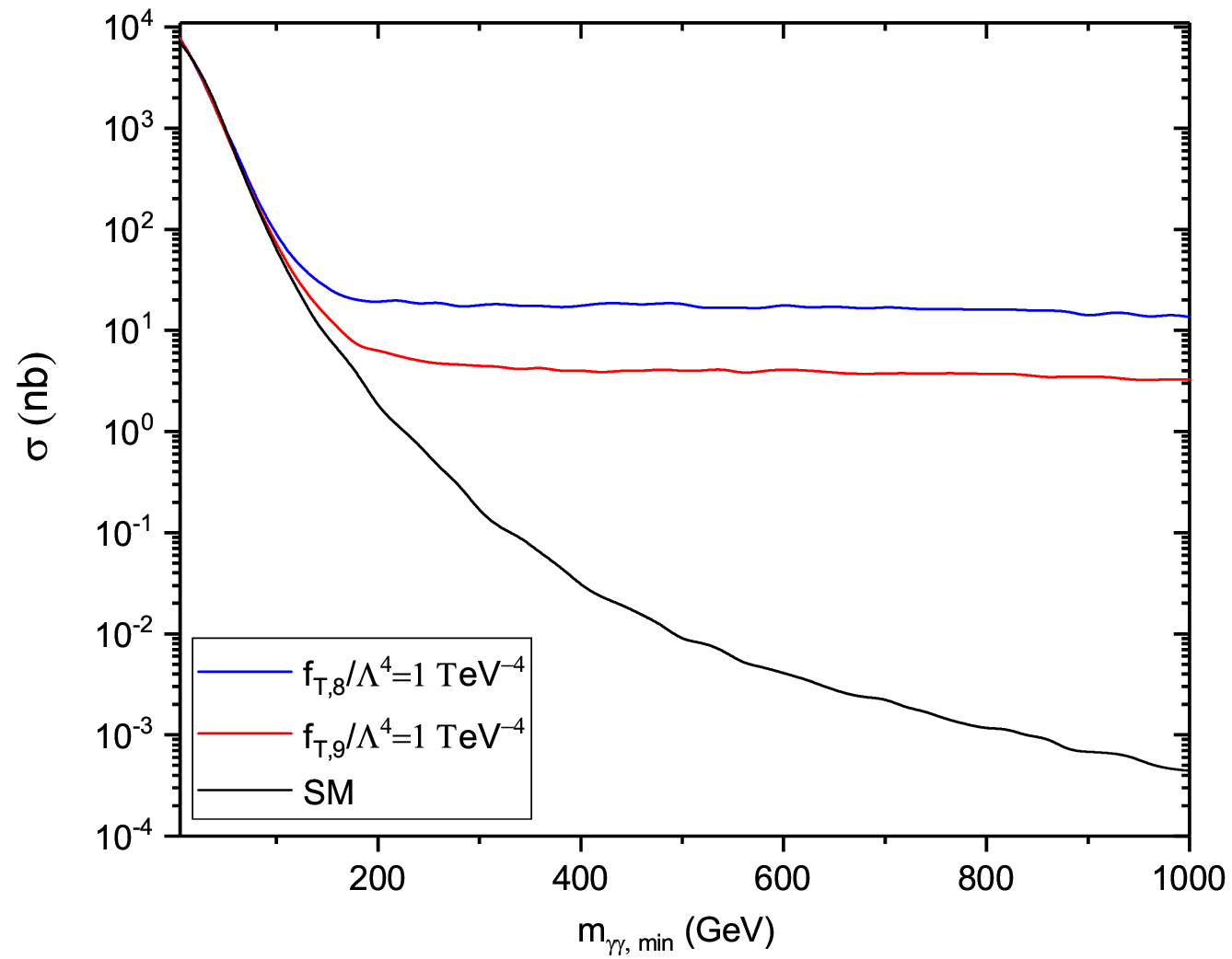}
\caption{The total cross section of the $\mathrm{Pb} \,\mathrm{Pb}
\rightarrow \mathrm{Pb}\, \gamma\gamma \,\mathrm{Pb}$ collision at
the FCC-hh as a function of the minimal value of the diphoton
invariant mass. $f_{T,8}/\Lambda^4$ ($f_{T,9}/\Lambda^4$) is fixed
to be 1 TeV$^{-4}$, with other aQGCs equal to zero. The SM cross
section is also shown (the lowest curve).}
\label{fig:CSMCUT}
\end{center}
\end{figure}

Let $S$ and $B$ be the total number of signal and SM events, and
$\delta$ is the percentage systematic uncertainty. The exclusion
significance is defined as
follows~\cite{Cowan:2011}-\cite{Zhang:2020}
\begin{align}\label{S_excl}
SS_{\mathrm{excl}} = \bigg\{ &2 \left[ S - B \ln\!\left( \frac{B + S
+ x}{2B} \right) - \frac{1}{\delta^2}\ln\!\left( \frac{B - S +
x}{2B}
\right) \right] \nonumber \\
&- (B + S - x) \left( 1 + \frac{1}{\delta^2B} \right) \bigg\}^{1/2}
,
\end{align}
with
\begin{equation}\label{x}
x = \sqrt{ (S + B)^2 - \frac{4S\delta^2B^2}{(1 + \delta^2 B)} } \;.
\end{equation}
We define the regions $SS_{\mathrm{excl}} \leqslant 1.645$ as
regions that can be excluded at the 95\% C.L. The discovery
significance is as follows \cite{Cowan:2011}-\cite{Zhang:2020}
\begin{equation}\label{S_disc}
SS_{\mathrm{disc}} = \sqrt{ 2\left[ (S + B) \ln\!\left( \frac{(B +
S)(1 + \delta^2 B)}{B + \delta^2 B(S + B)} \right) \right] -
\frac{1}{\delta^2}\ln\!\left( 1 + \frac{\delta^2 S}{1 + \delta^2 B}
\right) } \;.
\end{equation}
We classify the region with $SS_{\mathrm{disc}} > 5$ as discoverable
region at 5\,$\sigma$. In the limit $\delta \rightarrow 0$ we get
\begin{align}\label{SS_limit}
SS_{\mathrm{excl}} &= \sqrt{2[(S - B) \ln(1 + S/B)]} \;, \nonumber \\
SS_{\mathrm{disc}} &= \sqrt{2[(S + B) \ln(1 + S/B) - S]} \;.
\end{align}

The number of events is obtained via exclusive
$\gamma\gamma$-production in Pb-Pb collisions as
\begin{equation}\label{event_number}
\varepsilon_{\gamma\gamma} \cdot \sigma_{\gamma\gamma} \cdot
\mathcal{L}_{\mathrm{int}} \;,
\end{equation}
where $\sigma_{\gamma\gamma}$ is an exclusive cross section, and
$\mathcal{L}_{\mathrm{int}}$ is an integrated luminosity. The
combined signal efficiency in \eqref{event_number} is defined as
$\varepsilon_{\gamma\gamma} = \varepsilon_{\mathrm{acc}} \cdot
\varepsilon_{\mathrm{rec. id}}^2$, where
$\varepsilon_{\mathrm{acc}}$ takes into account trigger and
acceptance requirements, and $\varepsilon_{\mathrm{rec. id}}
\thickapprox 0.8$ is the photon reconstruction and identification
efficiency in the energy range of interest
\cite{Mangano:2017,Enterria:2013}. For the cuts we used (both
photons with $p_\bot > 5$ GeV within $|\eta^\gamma| < 2.5$),
$\varepsilon_{\mathrm{acc}} \thickapprox 0.4$ that results in
$\varepsilon_{\gamma\gamma} \thickapprox 0.26$.

\begin{figure}[htb]
\begin{center}
\includegraphics[scale=0.52]{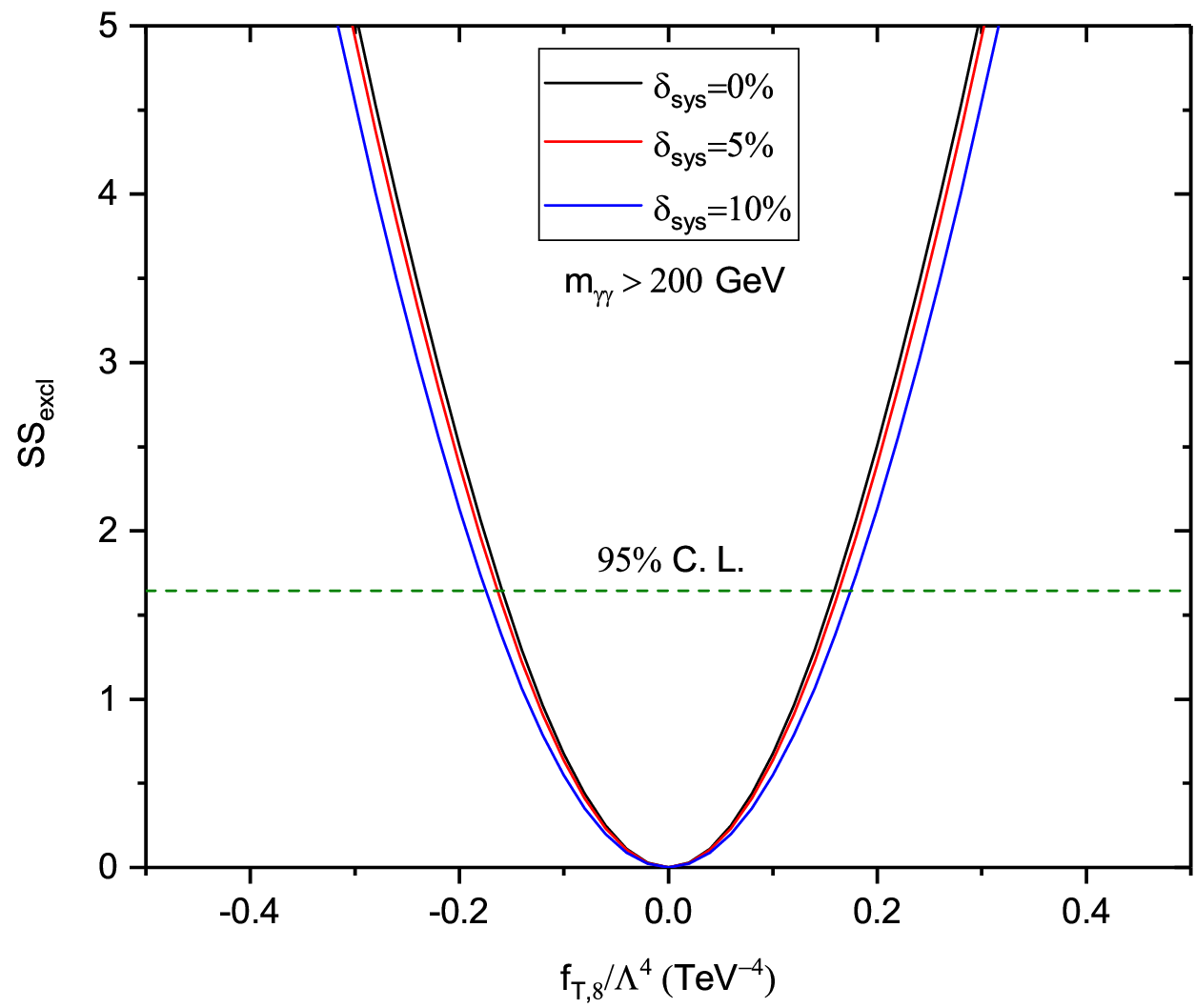}
\caption{The exclusion significance $SS_{\mathrm{excl}}$ as a
function of the coupling $f_{T,8}/\Lambda^4$ with the systematic
uncertainty $\delta_{\mathrm{sys}}$ in PbPb collision at the FCC-hh.
The cut on the invariant mass of the diphoton pair,
$m_{\gamma\gamma} > 200$ GeV, is used.}
\label{fig:SSexclw200ft8}
\end{center}
\end{figure}

\begin{figure}[htb]
\begin{center}
\includegraphics[scale=0.52]{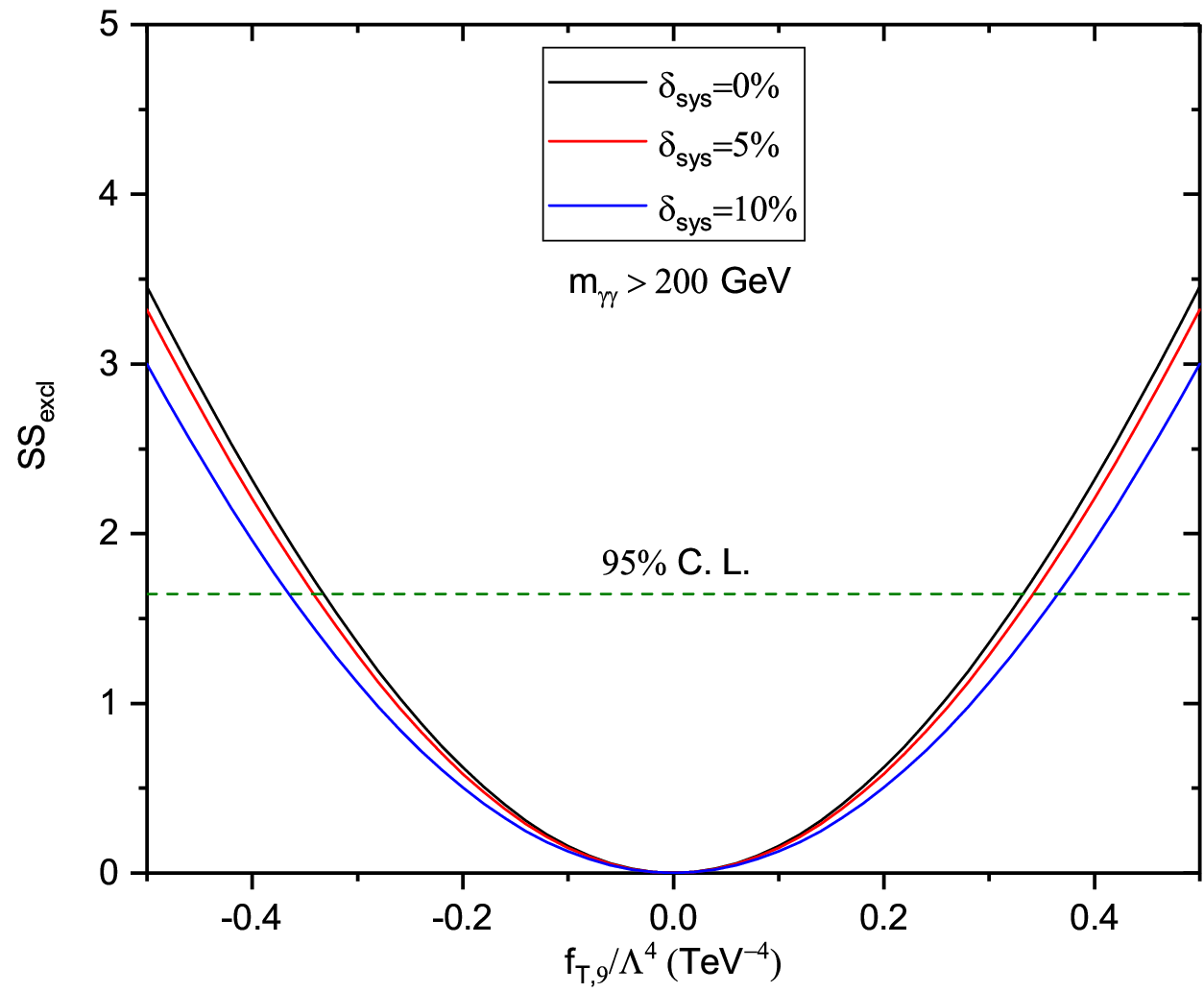}
\caption{The same as in Fig.~\ref{fig:SSexclw200ft8}, but for the
coupling $f_{T,9}/\Lambda^4$.}
\label{fig:SSexclw200ft9}
\end{center}
\end{figure}

\begin{figure}[htb]
\begin{center}
\includegraphics[scale=0.52]{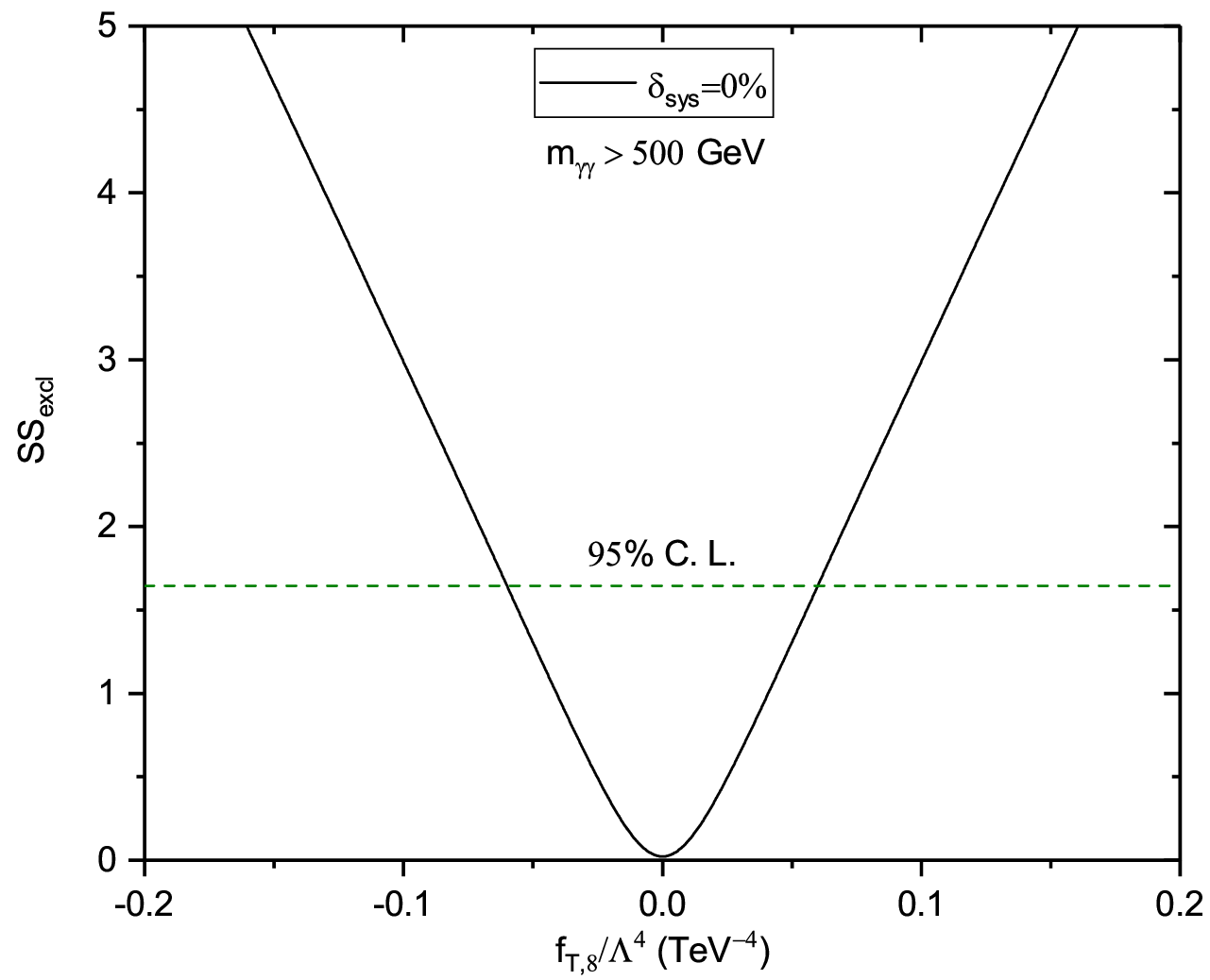}
\caption{The same as in Fig.~\ref{fig:SSexclw200ft8}, but for
$m_{\gamma\gamma} > 500$ GeV.}
\label{fig:SSexclw500ft8}
\end{center}
\end{figure}

\begin{figure}[htb]
\begin{center}
\includegraphics[scale=0.52]{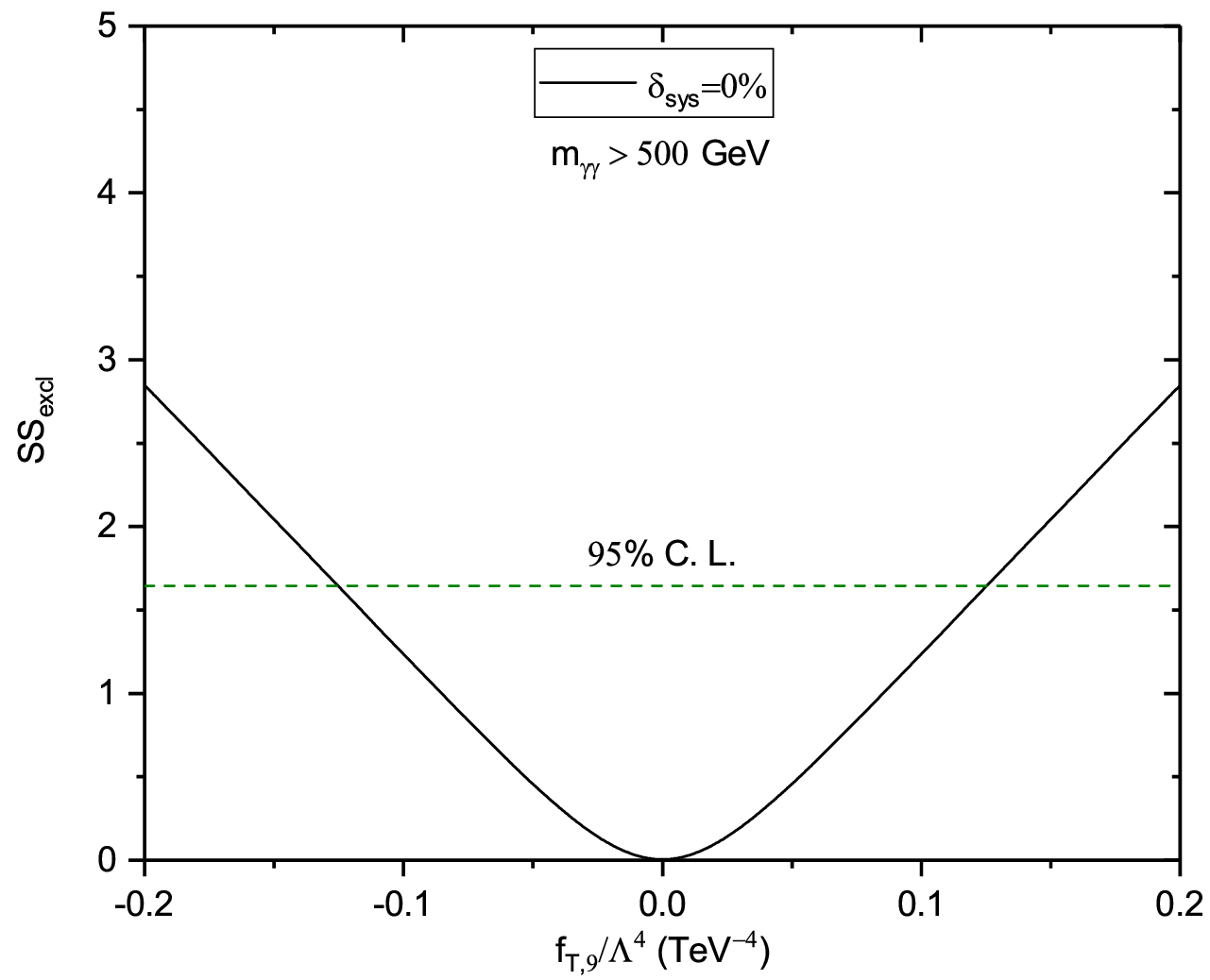}
\caption{The same as in Fig.~\ref{fig:SSexclw500ft8}, but for the
coupling $f_{T,9}/\Lambda^4$ and $m_{\gamma\gamma} > 500$ GeV.}
\label{fig:SSexclw500ft9}
\end{center}
\end{figure}

\begin{figure}[htb]
\begin{center}
\includegraphics[scale=0.52]{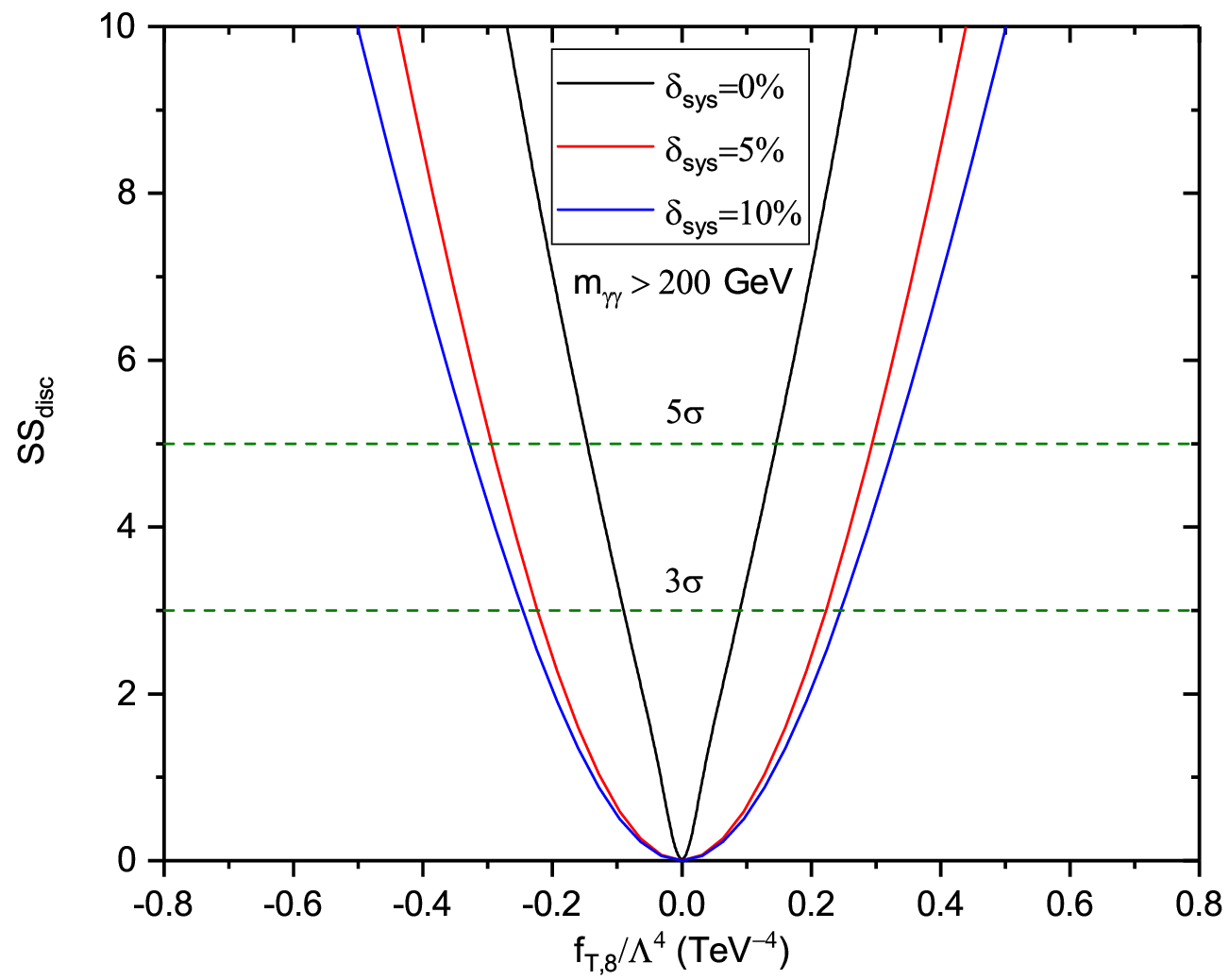}
\caption{The discovery significance $SS_{\mathrm{disc}}$ as a
function of the coupling $f_{T,8}/\Lambda^4$ with the systematic
uncertainty $\delta_{\mathrm{sys}}$ in PbPb collision at the FCC-hh.
The cut on the invariant mass of the diphoton pair,
$m_{\gamma\gamma} > 200$ GeV, is used.}
\label{fig:SSdisw200ft8}
\end{center}
\end{figure}

\begin{figure}[htb]
\begin{center}
\includegraphics[scale=0.52]{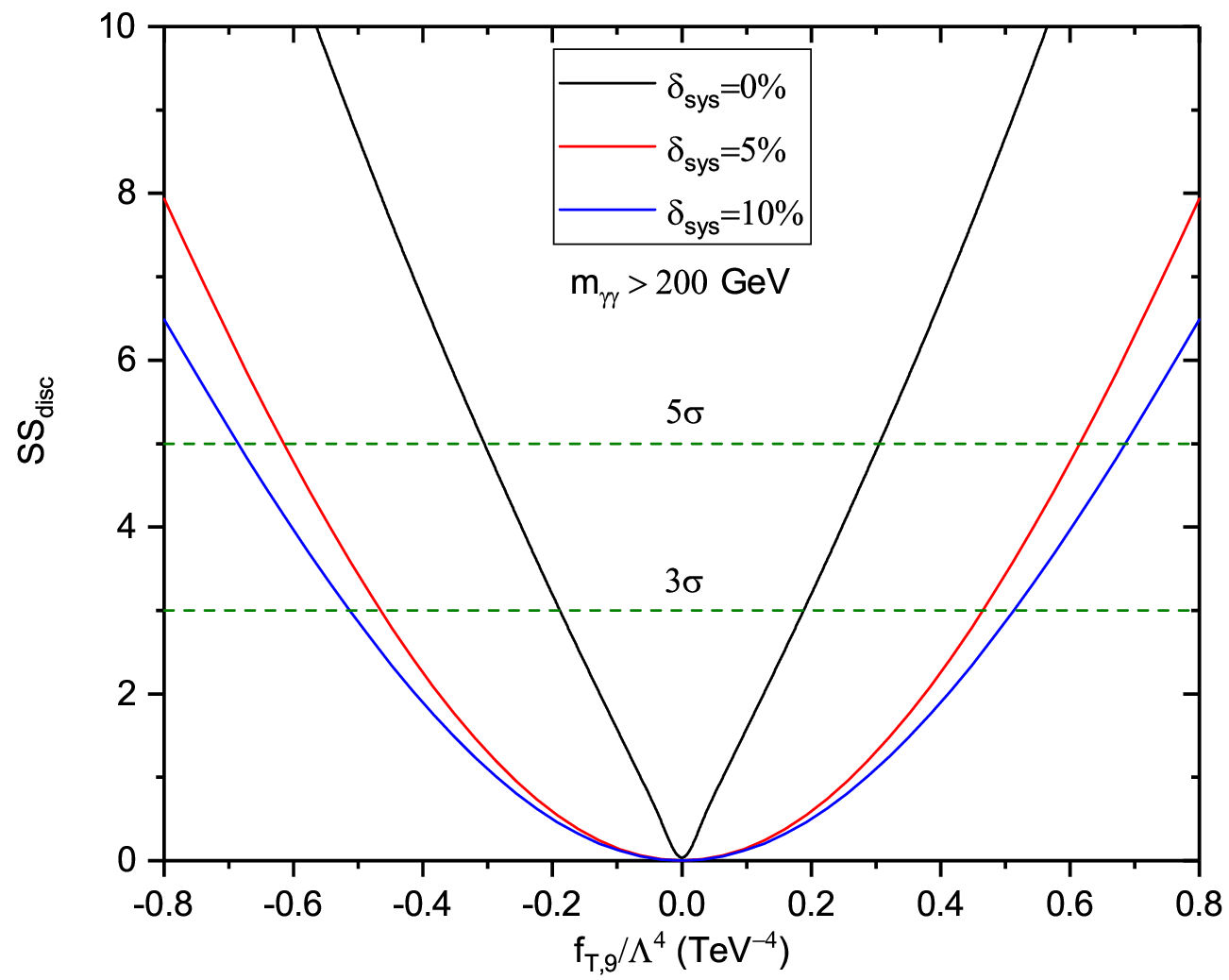}
\caption{The same as in Fig.~\ref{fig:SSdisw200ft8}, but for the
coupling $f_{T,9}/\Lambda^4$.}
\label{fig:SSdisw200ft9}
\end{center}
\end{figure}

\begin{figure}[htb]
\begin{center}
\includegraphics[scale=0.52]{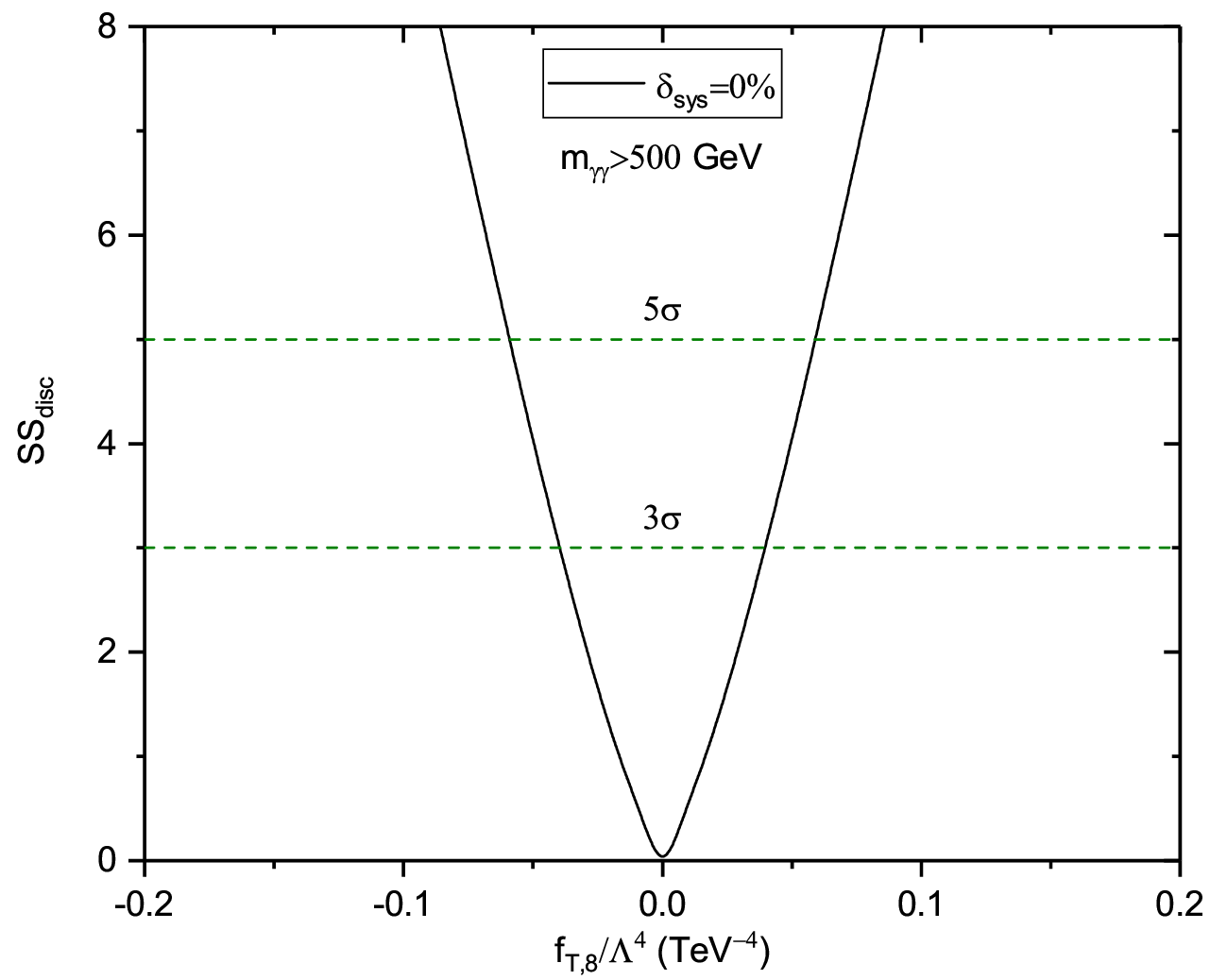}
\caption{The same as in Fig.~\ref{fig:SSdisw200ft8}, but for
$m_{\gamma\gamma} > 500$ GeV.}
\label{fig:SSdisw500ft8}
\end{center}
\end{figure}

\begin{figure}[htb]
\begin{center}
\includegraphics[scale=0.52]{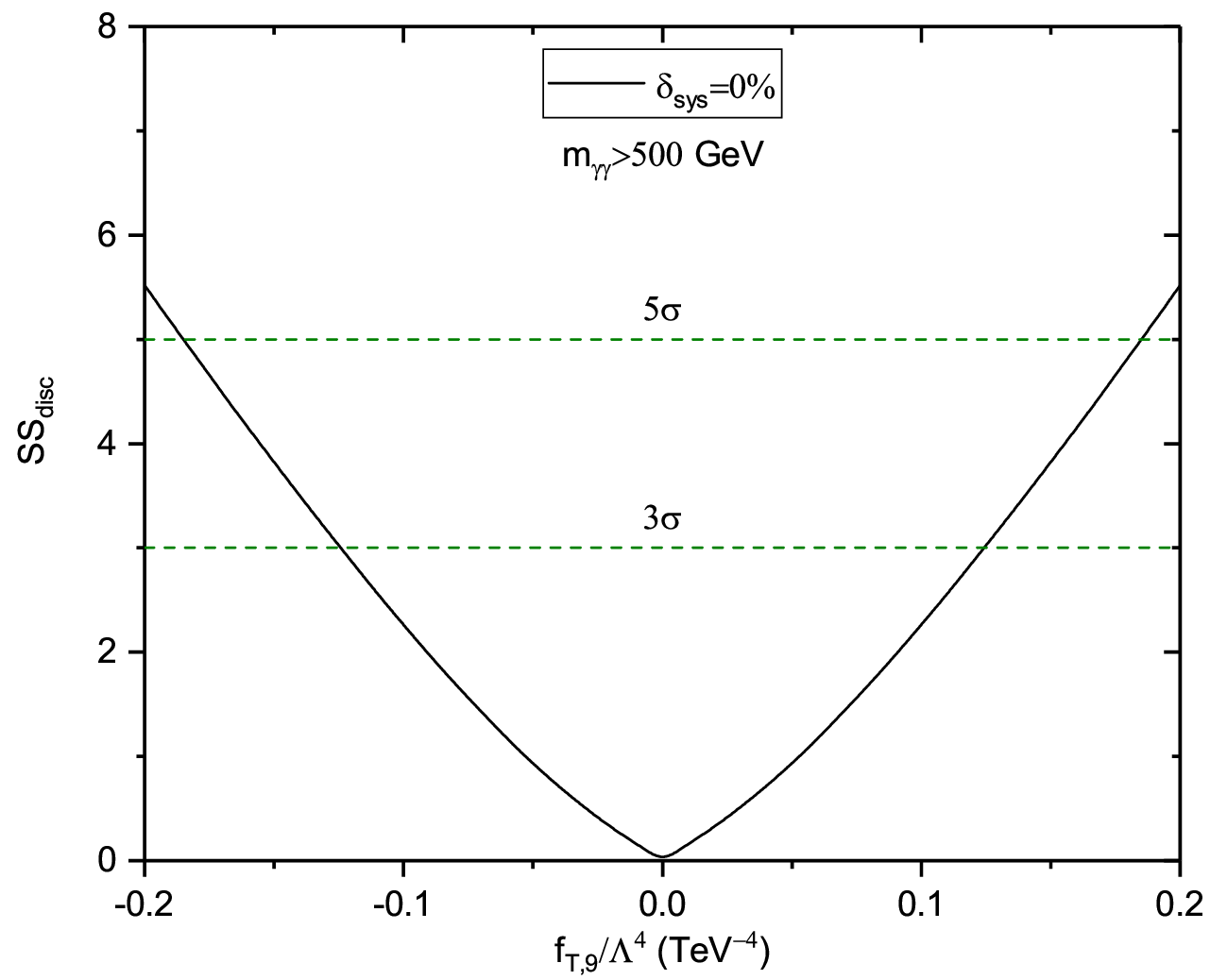}
\caption{The same as in Fig.~\ref{fig:SSdisw200ft8}, but for the
coupling $f_{T,9}/\Lambda^4$ and $m_{\gamma\gamma} > 500$ GeV.}
\label{fig:SSdisw500ft9}
\end{center}
\end{figure}

In Figs.~\ref{fig:SSexclw200ft8} and \ref{fig:SSexclw200ft9} we
present the exclusion significance $SS_{\mathrm{excl}}$ depending on
the couplings $f_{T,8}/\Lambda^4$ and $f_{T,9}/\Lambda^4$ when a cut
on the invariant mass of the final-state photons, $m_{\gamma\gamma}
> 200$ GeV, is applied. The curves correspond to three different
values of a systematic uncertainty $\delta_{\mathrm{sys}}$. Next two
figures, \ref{fig:SSexclw500ft8} and \ref{fig:SSexclw500ft9},
demonstrate $SS_{\mathrm{excl}}$ when the stronger cut
$m_{\gamma\gamma} > 500$ GeV is imposed. Note that in such a case,
the exclusion significance depends very weakly on
$\delta_{\mathrm{sys}}$ (at least, for $\delta_{\mathrm{sys}} \leq
10\%$). That is why we present our results for
$\delta_{\mathrm{sys}} = 0$ only.
%
%%%%%%%%%%%%%%%%%%%%%%%%%%%%%%%%%%%%%%%%%%%%%%%%
% Figure 3. Exclusion significance fT8 200 GeV %
%%%%%%%%%%%%%%%%%%%%%%%%%%%%%%%%%%%%%%%%%%%%%%%%
%
%%%%%%%%%%%%%%%%%%%%%%%%%%%%%%%%%%%%%%%%%%%%%%%%
% Figure 4. Exclusion significance fT9 200 GeV %
%%%%%%%%%%%%%%%%%%%%%%%%%%%%%%%%%%%%%%%%%%%%%%%%
%
%%%%%%%%%%%%%%%%%%%%%%%%%%%%%%%%%%%%%%%%%%%%%%%%
% Figure 5. Exclusion significance fT8 500 GeV %
%%%%%%%%%%%%%%%%%%%%%%%%%%%%%%%%%%%%%%%%%%%%%%%%
%
%%%%%%%%%%%%%%%%%%%%%%%%%%%%%%%%%%%%%%%%%%%%%%%%
% Figure 6. Exclusion significance fT9 500 GeV %
%%%%%%%%%%%%%%%%%%%%%%%%%%%%%%%%%%%%%%%%%%%%%%%%
The discovery significance $SS_{\mathrm{disc}}$ are shown in
Figs.~\ref{fig:SSdisw200ft8}-\ref{fig:SSdisw500ft9}. Note, for the
cut $m_{\gamma\gamma} > 500$ GeV  there is no notable dependence on
the systematic uncertainty $\delta_{\mathrm{sys}}$ (see
Figs.~\ref{fig:SSdisw500ft8}, \ref{fig:SSdisw500ft9}), as it does
for the exclusion significance $SS_{\mathrm{excl}}$.
%
%%%%%%%%%%%%%%%%%%%%%%%%%%%%%%%%%%%%%%%%%%%%%%%%
% Figure 7. Discovery significance fT8 200 GeV %
%%%%%%%%%%%%%%%%%%%%%%%%%%%%%%%%%%%%%%%%%%%%%%%%
%
%%%%%%%%%%%%%%%%%%%%%%%%%%%%%%%%%%%%%%%%%%%%%%%%
% Figure 8. Discovery significance fT9 200 GeV %
%%%%%%%%%%%%%%%%%%%%%%%%%%%%%%%%%%%%%%%%%%%%%%%%
%
%%%%%%%%%%%%%%%%%%%%%%%%%%%%%%%%%%%%%%%%%%%%%%%%
% Figure 9. Discovery significance fT8 500 GeV %
%%%%%%%%%%%%%%%%%%%%%%%%%%%%%%%%%%%%%%%%%%%%%%%%
%
%%%%%%%%%%%%%%%%%%%%%%%%%%%%%%%%%%%%%%%%%%%%%%%%%
% Figure 10. Discovery significance fT9 500 GeV %
%%%%%%%%%%%%%%%%%%%%%%%%%%%%%%%%%%%%%%%%%%%%%%%%%
Our predictions for all couplings $f_{T,i}/\Lambda^4$ ($i =
0,1,2,5,6,7,8,9$) are collected in Tabs.~\ref{tab:1}, \ref{tab:2}.
As one can see, the best bounds are obtained for the couplings
$f_{T,8}/\Lambda^4$ and $f_{T,9}/\Lambda^4$. That is because in
eq.~\eqref{g1_g2_vs_fTi} they are multiplied by the factor $c_w^4$,
while other couplings are multiplied by powers of a small quantity
$s_w$. As a result, the cross section appears to be more sensitive
to the couplings $f_{T,8}/\Lambda^4$ and $f_{T,9}/\Lambda^4$. All
bounds become stronger if the cut $m_{\gamma\gamma} > 500$ GeV is
used instead of the cut $m_{\gamma\gamma} > 200$ GeV.

%%%%%%%%%%%%%%%%%%%%%%%%%
% Bounds on fTi 200 GeV %
%%%%%%%%%%%%%%%%%%%%%%%%%
\begin{table}
\begin{center}
\begin{tabular}{|c||c||c|c|}
  \hline \hline
  aQGC & $\delta_{\mathrm{sys}}$ & 95\% C.L. & $5\sigma$ \\
  \hline\hline
     & 0\% & $[-1.91,1.91]$ & $[-1.77,1.77]$ \\
$f_{T,0}/\Lambda^4$, $f_{T,1}/\Lambda^4$ & 5\% & $[-2.01,2.01]$ & $[-2.52,2.52]$ \\
     & 10\% & $[-2.17,2.17]$ & $[-3.52,3.52]$ \\
  \hline
     & 0\% & $[-3.75,3.75]$ & $[-3.64,3.64]$ \\
$f_{T,2}/\Lambda^4$ & 5\% & $[-4.15,4.15]$ & $[-7.40,7.40]$ \\
     & 10\% & $[-4.49,4.49]$ & $[-8.37,8.37]$ \\
  \hline
     & 0\% & $[-0.55,0.55]$ & $[-0.51,0.51]$ \\
$f_{T,5}/\Lambda^4$, $f_{T,6}/\Lambda^4$ & 5\% & $[-0.58,0.58]$ & $[-1.01,1.01]$ \\
     & 10\% & $[-0.62,0.62]$ & $[-1.14,1.14]$ \\
  \hline
     & 0\% & $[-1.07,1.07]$ & $[-1.04,1.04]$ \\
$f_{T,7}/\Lambda^4$ & 5\% & $[-1.19,1.19]$ & $[-2.12, 2.12]$ \\
     & 10\% & $[-1.29, 1.29]$ & $[-2.40, 2.40]$ \\
  \hline
     & 0\% & $[-0.16, 0.16]$ & $[-0.15, 0.15]$ \\
$f_{T,8}/\Lambda^4$ & 5\% & $[-0.17, 0.17]$ & $[-0.29, 0.29]$ \\
     & 10\% & $[-0.24, 0.24]$ & $[-0.33, 0.33]$ \\
  \hline
     & 0\% & $[-0.31, 0.31]$ & $[-0.30, 0.30]$ \\
$f_{T,9}/\Lambda^4$ & 5\% & $[-0.35, 0.35]$ & $[-0.61, 0.61]$ \\
     & 10\% & $[-0.37, 0.37]$ & $[-0.68, 0.68]$ \\
  \hline\hline
\end{tabular}
\end{center}
\caption{The bounds on aQGCs for the $\gamma\gamma$ production in
$\mathrm{Pb Pb}$ collisions at the 100 TeV FCC-hh, with the cut on
the diphoton invariant mass $m_{\gamma\gamma} > 200$ GeV. The
couplings are in units of [TeV]$^{-4}$. The integrated luminosity is
taken to be 110 nb$^{-1}$ \protect\cite{Yang:2024,Jiang:2024}.}
\label{tab:1}
\end{table}

%%%%%%%%%%%%%%%%%%%%%%%%%
% Bounds on fTi 500 GeV %
%%%%%%%%%%%%%%%%%%%%%%%%%
\begin{table}
\begin{center}
\begin{tabular}{|c||c||c|c|}
  \hline \hline
  aQGC & $\delta_{\mathrm{sys}}$ & 95\% C.L. & $5\sigma$ \\
  \hline\hline
$f_{T,0}/\Lambda^4$, $f_{T,1}/\Lambda^4$  & 0\% & $[-0.73, 0.73]$ & $[-1.07,1.07]$ \\
  \hline
$f_{T,2}/\Lambda^4$ & 0\% & $[-1.53, 1.53]$ & $[-2.21, 2.21]$
\\
  \hline
$f_{T,5}/\Lambda^4$, $f_{T,6}/\Lambda^4$ & 0\% & $[-0.21, 0.21]$ & $[-0.30, 0.30]$ \\
  \hline
$f_{T,7}/\Lambda^4$ & 0\% & $[-0.44,0.44]$ & $[-1.04, 1.04]$ \\
  \hline
$f_{T,8}/\Lambda^4$ & 0\% & $[-0.060,0.060]$ & $[-0.088,0.088]$ \\
  \hline
$f_{T,9}/\Lambda^4$ & 0\% & $[-0.13, 0.13]$ & $[-0.180,0.180]$ \\
  \hline\hline
\end{tabular}
\end{center}
\caption{The same as in Tab.~\ref{tab:1}, but when the cut
$m_{\gamma\gamma} > 500$ GeV is used.}
\label{tab:2}
\end{table}

In our previous papers \cite{I_K:2021_1,I_K:2021_2} the following
unitarity constraints have been derived
\begin{align}\label{unitarity_lim_anal}
|g_1| &< \frac{32\pi}{\hat{s}^2} = \frac{2\pi}{\tau^2} \ (g_2 = 0) \;,
\nonumber \\
|g_2| &< \frac{16\pi}{\hat{s}^2} = \frac{\pi}{\tau^2}  \ (g_1 = 0)
\;,
\end{align}
where $\hat{s} = 4 \tau$ is the center-off-mass energy squared of
the LbL process. As shown in Appendix~C, although $\tau$ can vary up
to $(E_N - m_N)^2$, a main contribution to the cross section comes
from the region $\tau \approx \tau_{\mathrm{eff}}$, where
$\sqrt{\tau_{\mathrm{eff}}}$ is about one-few TeV. Then the
unitarity upper bounds on the couplings $f_{T,i}$ can be estimated
as $(0.2\div 3.0) \mathrm{\ TeV}^{-4}$. It allows us to conclude
that the unitarity is not violated in the region of aQGCs considered
in the present paper.

%%%%%%%%%%%%%%%%%%%%%%%
\section{Conclusions} %
%%%%%%%%%%%%%%%%%%%%%%%

By considering the diphoton production in lead-lead collisions at
the 100 TeV FCC-hh, we have obtained the $5\sigma$ and 95\% C.L.
bounds on the anomalous quartic gauge couplings, depending on the
cut on photon-pair invariant mass and possible systematic
uncertainties. They are presented in figures
\ref{fig:SSexclw200ft8}-\ref{fig:SSdisw500ft9} and tables
\ref{tab:1} and \ref{tab:2}.

Let us compare our results with the current experimental constraints
on aQGCs and predictions of other authors. The observed
one-dimensional limits on dimension-8 aQGCs obtained recently by the
ATLAS collaboration are the following \cite{ATLAS:QGCs_5}:
\begin{align}\label{fT_ATLAS}
f_{T,0}/\Lambda^4: &\quad [-0.25, 0.22] \mathrm{\ TeV}^{-4}
\;, \nonumber \\
f_{T,1}/\Lambda^4: &\quad [-0.24, 0.24] \mathrm{\ TeV}^{-4}
\;, \nonumber \\
f_{T,2}/\Lambda^4: &\quad [-0.55, 0.55] \mathrm{\ TeV}^{-4}
\;, \nonumber \\
f_{T,5}/\Lambda^4: &\quad [-0.64, 0.58] \mathrm{\ TeV}^{-4}
\;, \nonumber \\
f_{T,6}/\Lambda^4: &\quad [-0.74, 0.71] \mathrm{\ TeV}^{-4}
\;, \nonumber \\
f_{T,7}/\Lambda^4: &\quad [-1.94, 1.70] \mathrm{\ TeV}^{-4}
\;, \nonumber \\
f_{T,8}/\Lambda^4: &\quad [-0.48, 0.48] \mathrm{\ TeV}^{-4}
\;, \nonumber \\
f_{T,9}/\Lambda^4: &\quad [-1.02, 1.03] \mathrm{\ TeV}^{-4} \;.
\end{align}
The limits are obtained by setting a single operator at a time.
These constraints are either comparable with or more stringent than
those obtained previously by the CMS collaboration
\cite{CMS:QGCs_4}:
\begin{align}\label{fT_ATLAS}
f_{T,0}/\Lambda^4: &\quad [-0.47, 0.51] \mathrm{\ TeV}^{-4}
\;, \nonumber \\
f_{T,1}/\Lambda^4: &\quad [-0.31, 0.34] \mathrm{\ TeV}^{-4}
\;, \nonumber \\
f_{T,2}/\Lambda^4: &\quad [-0.85, 0.10] \mathrm{\ TeV}^{-4}
\;, \nonumber \\
f_{T,5}/\Lambda^4: &\quad [-0.31, 0.33] \mathrm{\ TeV}^{-4}
\;, \nonumber \\
f_{T,6}/\Lambda^4: &\quad [-0.25, 0.27] \mathrm{\ TeV}^{-4}
\;, \nonumber \\
f_{T,7}/\Lambda^4: &\quad [-0.67, 0.73] \mathrm{\ TeV}^{-4} \;.
\end{align}
As one can see, for a number of couplings (namely:
$f_{T,5}/\Lambda^4, f_{T,6}/\Lambda^4, f_{T,7}/\Lambda^4$, and
$f_{T,9}/\Lambda^4$) our limits are more stringent than the current
LHC limits.

Recently, bounds on the couplings $f_{T,8}/\Lambda^4$ and
$f_{T,9}/\Lambda^4$ at $3\sigma$, $5\sigma$, and 95\% C.L. for the
integrated luminosity of $\mathcal{L} = 30$ ab$^{-1}$ in $pp$
collisions at the FCC-hh were reported in \cite{Senol:2025}. The
best bound on $f_{T,8}/\Lambda^4$ is expected to be $[-4.84,
4.84]\times10^{-3}$ TeV$^{-4}$, while for $f_{T,9}/\Lambda^4$ it is
equal to $[-2.46, 2.46]\times10^{-2}$ TeV$^{-4}$~\cite{Senol:2025}.

In another recent paper \cite{Chen:2024}, a search of aQGCs in the
$\mu^+\mu^- \rightarrow \bar{\nu}\nu \gamma\gamma$ process was
presented. The expected coefficient constraints calculated by the
local outlier factor (LOF) method is (for the collision energy of
10 TeV and integrated luminosity of 10 ab$^{-1}$) \cite{Chen:2024}:
\begin{align}\label{fT_muon_collider}
f_{T,0}/\Lambda^4: &\quad [-1.32, 0.051] \times 10^{-3} \mathrm{\
TeV}^{-4}
\;, \nonumber \\
f_{T,1}/\Lambda^4: &\quad [-1.82, 2.22] \ \,\times 10^{-3} \mathrm{\
TeV}^{-4}
\;, \nonumber \\
f_{T,2}/\Lambda^4: &\quad [-6.75, 0.80] \ \,\times 10^{-3} \mathrm{\
TeV}^{-4}
\;, \nonumber \\
f_{T,5}/\Lambda^4: &\quad [-0.40, 0.027] \times 10^{-3} \mathrm{\
TeV}^{-4}
\;, \nonumber \\
f_{T,6}/\Lambda^4: &\quad [-0.70, 0.066] \times 10^{-3} \mathrm{\
TeV}^{-4}
\;, \nonumber \\
f_{T,7}/\Lambda^4: &\quad [-1.89, 0.060] \times 10^{-3} \mathrm{\
TeV}^{-4} \;.
\end{align}
The expected sensitivity of the 10 TeV muon collider with the
integrated luminosity of 10 ab$^{-1}$ on the anomalous couplings
$f_{T,i}/\Lambda^4$ was also studied in \cite{Gutierrez:2025}
through the $\mu^+\mu^- \rightarrow \mu^+\mu^- Z\gamma$ signal. The
constraints of the order of $\mathrm{O}(10^{-3}- 10^{-2})$
TeV$^{-4}$ were obtained \cite{Gutierrez:2025}.

The bounds \eqref{fT_muon_collider} can be compared with our
constraints previously obtained for a future muon collider with the
center-off-mass energy of 3 TeV, 14 TeV, and 100 TeV, and integrated
luminosity of 1 ab$^{-1}$, 20 ab$^{-1}$, and 1000 ab$^{-1}$,
respectively \cite{I_K:2024,I_K:2025}.

%%%%%%%%%%%%%%%%%%%%%%%%%%%%%%%%%%%%%%%%%%%%%%%%%%%%%%%%%%%%%%%%%%%%%

%%%%%%%%%%%%%%
% Appendix A %
%%%%%%%%%%%%%%

\setcounter{equation}{0}
\renewcommand{\theequation}{A.\arabic{equation}}

\setcounter{section}{0}
\renewcommand{\thesection}{A.\arabic{section}}

%%%%%%%%%%%%%%%%%%%%%%%%%%%%%%%%%%%%%%%%%%%%%%%%%%%%%%
\section*{Appendix A. Anomalous helicity amplitudes} %
%%%%%%%%%%%%%%%%%%%%%%%%%%%%%%%%%%%%%%%%%%%%%%%%%%%%%%

For the LbL scattering the Bose-Einstein statistics and parity
invariance demand that there exist six independent anomalous
helicity amplitudes $M_{\lambda_1\lambda_2\lambda_3\lambda_4}$ with
$\lambda_1 = +1$. Only three of them are nonzero \cite{I_K:2024},
\begin{align}\label{independent_helicity_ampl_gamma}
M_{++++}(s,t,u) &= \frac{(4g_1 + 3g_2)}{2} \,s^2 \;,
\nonumber \\
M_{++--}(s,t,u) &=  (4g_1 + g_2) \,(t^2 + tu + u^2)\;,
\nonumber \\
M_{+-+-}(s,t,u) &=  \frac{(4g_1 + 3g_2)}{2} \,u^2 \;,
\end{align}
where $s$, $t$, $u$ are Mandelstam variables of the
$\gamma\gamma\rightarrow\gamma\gamma$ collision, $s+t+u = 0$. Other
three independent helicity amplitudes with $\lambda_1 = +1$ are
equal to zero,
\begin{equation}\label{zero_helicity_ampl_gamma}
M_{+++-} = M_{++-+} = M_{+-++} = 0 \;.
\end{equation}
We also have the crossing relations
\begin{align}\label{dependent_helicity_ampl_gamma}
M_{+--+}(s,t,u) &= M_{+-+-}(s,u,t) = \frac{(4g_1 + 3g_2)}{2} \,t^2
\;,
\nonumber \\
M_{+---}(s,t,u) &= M_{+-++}(s,u,t) = 0 \;.
\end{align}
Equations
\eqref{independent_helicity_ampl_gamma}-\eqref{dependent_helicity_ampl_gamma}
represent eight helicity amplitudes with $\lambda_1 = +1$. The
helicity amplitudes with $\lambda_1 = -1$ can be obtained from them
by the use of the parity relation,
\begin{equation}\label{parity_relations}
M_{-\lambda_2\lambda_3\lambda_4}(s,t,u) =
M_{+-\lambda_2-\lambda_3-\lambda_4}(s,t,u) \;.
\end{equation}

%%%%%%%%%%%%%%
% Appendix B %
%%%%%%%%%%%%%%

\setcounter{equation}{0}
\renewcommand{\theequation}{B.\arabic{equation}}

%%%%%%%%%%%%%%%%%%%%%%%%%%%%%%%%%%%%%%%%%%%%%%%%%
\section*{Appendix B. Two-photon cross section} %
%%%%%%%%%%%%%%%%%%%%%%%%%%%%%%%%%%%%%%%%%%%%%%%%%

The exclusive cross section for two-photon process can be
schematically written as
\begin{equation}\label{two-photon_cs}
d\sigma = \int \!d\omega_1 \!\int \!d\omega_2 \frac{dn}{d\omega_1}
\frac{dn}{d\omega_2} \,d\hat{\sigma} \;,
\end{equation}
where the photon flux from a charge $Z$ nucleus is of the form
\cite{Baltz:2008}
\begin{align}\label{photon_flux}
\frac{dn}{d\omega} = \frac{2Z^2\alpha}{\pi \omega} & \bigg\{
\left(\frac{\omega}{\omega_0}\right)
K_0\left(\frac{\omega}{\omega_0}\right)
K_1\left(\frac{\omega}{\omega_0}\right) \nonumber \\
&- \left(\frac{\omega}{\omega_0}\right)^{\!\!2} \!\!\left[
K_1^2\left(\frac{\omega}{\omega_0}\right) -
K_0^2\left(\frac{\omega}{\omega_0}\right) \right] \bigg\} ,
\end{align}
with $\omega_0 = E_N/(m_N R_A) = \gamma_L/R_A$. The cross section
\eqref{two-photon_cs} can be rewritten in the form
\begin{equation}\label{two-photon_cs_m}
d\sigma = \int \frac{d\omega_1}{\omega_1} \!\int
\frac{d\omega_2}{\omega_2} f(\omega_1) f(\omega_2) \,d\hat{\sigma}
\;,
\end{equation}
where the photon distribution in the nucleus $f(\omega)$ is defined
by eq.~\eqref{dist_gamma_nucleus}.%
\footnote{As one can see, $dn/d\omega$ and $f(\omega)$ differ by the
factor $1/\omega$.}
After change of variables, $\omega = \omega_1$, $\tau =
\omega_1\omega_2$, we come to formula \eqref{cs}.

Let us define an energy fraction of the photon, $x = \omega/E_N$.
For small $x$, the photon flux is given by
\begin{equation}\label{pgoton_flux_small_x}
\frac{dn_{\gamma/N}(x)}{dx}\bigg|_{x \ll 1} \simeq
\frac{Z^2\alpha}{\pi x} \ln \frac{1}{(x m_N R_A)^2} \;.
\end{equation}
If the source of the emitted photon is instead an electron, then
\begin{equation}\label{replacements}
Z \rightarrow 1, \quad m_N \rightarrow m_e, \quad R_A \rightarrow
1/Q \;,
\end{equation}
that would give the flux
\begin{equation}\label{pgoton_flux_small_x}
\frac{dn_{\gamma/e}(x)}{dx}\bigg|_{x \ll 1} \simeq \frac{\alpha}{\pi
x} \ln \frac{Q^2}{x^2 m_e^2} \;,
\end{equation}
in agreement with the Weizs\"{a}cker-Williams form.

%%%%%%%%%%%%%%
% Appendix C %
%%%%%%%%%%%%%%

\setcounter{equation}{0}
\renewcommand{\theequation}{C.\arabic{equation}}

%%%%%%%%%%%%%%%%%%%%%%%%%%%%%%%%%%%%%%%%%%%%%%%%%%%%%%%%%%%%%%%%%%%%%%%%%%%%%%%%%%%%%%%%%%%%
\section*{Appendix C. Effective energy of $\gamma\gamma\rightarrow\gamma\gamma$ collision} %
%%%%%%%%%%%%%%%%%%%%%%%%%%%%%%%%%%%%%%%%%%%%%%%%%%%%%%%%%%%%%%%%%%%%%%%%%%%%%%%%%%%%%%%%%%%%

The photon luminosity as a function of variable $\tau$ is as follows
\begin{equation}\label{L_ph}
\mathcal{L}_{\mathrm{ph}}(\tau) =
\int\limits_{\omega_{\min}}^{\omega_{\max}}
\!\!\frac{d\omega}{\omega} f_{\gamma/N}(\omega)
f_{\gamma/N}(\tau/\omega) \;,
\end{equation}
where $\tau$ is in units of TeV$^2$. Remember that $\sqrt{\tau}$ is
equal to the half an invariant energy of the
$\gamma\gamma\rightarrow \gamma\gamma$ scattering. Note that the
anomalous cross section is proportional to $\tau^4$. Thus, taking
into account eqs.~\eqref{cs} and \eqref{L_ph}, we can define an
average value of $\tau$ as
\begin{equation}\label{tau_av_def}
\langle \tau \rangle = \frac{1}{N}
\int\limits_{\tau_{\min}}^{\tau_{\max}} d\tau \tau^4
\mathcal{L}_{\mathrm{ph}}(\tau) \;,
\end{equation}
where
\begin{equation}\label{norm}
N = \int\limits_{\tau_{\min}}^{\tau_{\max}} d\tau \tau^3
\mathcal{L}_{\mathrm{eff}}(\tau) \;,
\end{equation}
and $\tau_{\max}$, $\tau_{\min}$ are defined by
eqs.~\eqref{upper_limits}, \eqref{lower_limits}.

For the lead-lead collision at the 100 TeV FCC-hh our calculations
result in
\begin{equation}\label{tau_av_FCC-hh}
\langle \tau \rangle \simeq 1.23 \mathrm{\ TeV}^2 \;.
\end{equation}
It means that a main contribution to the cross section \eqref{cs}
comes from a relatively small $\tau$ region, $\tau \ll \tau_{\max} =
(E_N - m_N)^2$. The numerical calculations really demonstrate that
the cross section remains almost unchanged, if one takes the upper
limit in \eqref{cs} to be a few tens of TeV$^2$.

%%%%%%%%%%%%%%%%%%%%%%%%%%%%%%%%%%%%%%%%%%%%%%%%%%%%%%%%%%%%%%%%%%%%

%%%%%%%%%%%%%%
% References %
%%%%%%%%%%%%%%

%%%%%%%%%%%%%%%%%%%%%%%%%%%%%%%%%%%%%%%%%%%%%%%%%%%%%%%%%%%%%%%%%%%%%

%%%%%%%%%%%%%%%
\end{document}